\documentclass[sigconf, nonacm=True]{acmart}
\usepackage{array} 
\usepackage{graphicx}
\usepackage{subcaption}

\AtBeginDocument{%
  }

\setcopyright{rightsretained}
\copyrightyear{2025}
\acmYear{2025}
\begin{document}

\title{Material Identification Via RFID For Smart Shopping}


\author{David Wang}
\affiliation{%
  \institution{University of Michigan}
   \city{Ann Arbor}
   \state{Michigan}
   \country{USA}}
\email{davwan@umich.edu}

\author{Derek Goh}
\affiliation{%
  \institution{University of Michigan}
  \city{Ann Arbor}
  \state{Michigan}
  \country{USA}}
\email{derekgoh@umich.edu}

\author{Jiale Zhang}
\affiliation{%
  \institution{University of Michigan}
  \city{Ann Arbor}
  \state{Michigan}
  \country{USA}}
\email{jiale@umich.edu}
\renewcommand{\shortauthors}{Wang et al.}

\begin{abstract}
Cashierless stores rely on computer vision and RFID tags to associate shoppers with items, but concealed items placed in backpacks, pockets, or bags create challenges for theft prevention. We introduce a system that turns existing RFID tagged items into material sensors by exploiting how different containers attenuate and scatter RF signals. Using RSSI and phase angle, we trained a neural network to classify seven common containers. In a simulated retail environment, the model achieves 89\% accuracy with one second samples and 74\% accuracy from single reads. Incorporating distance measurements, our system achieves 82\% accuracy across 0.3–2m tag to reader separations. When deployed at aisle or doorway choke points, the system can flag suspicious events in real time, prompting camera screening or staff intervention. By combining material identification with computer vision tracking, our system provides proactive loss prevention for cashierless retail while utilizing existing infrastructure.
\end{abstract}




\maketitle

\section{Introduction}
Cashierless stores are a promising trend that aim to create a more streamlined and user-friendly retail experience. Compared to traditional stores, cashierless stores offer a faster "just walk out" checkout experience for consumers and serve as an accurate way for store owners to manage inventory \cite{kwon2023effects, schogel2020cashierless}. Cashierless stores incorporate a suite of sensors to track customer activity and accurately associate shoppers with the items they take. These often include computer vision, RFID, vibration sensing, weight sensing, and vibration detection, typically combined into a multi-modal system with sensor fusion \cite{ruiz2019aim3s, rukundo2025survey, szabo2023systematic, low2021developing}. A common cashierless store implementation uses RFID tags for item tracking, as they can be wirelessly scanned at checkout \cite{bocanegra2020rfgo}, and computer vision for customer movement tracking. However, a major challenge with this system is theft prevention, an essential part of retail systems. Shoppers may place items inside shopping bags or boxes, while thieves may intentionally try to hide items inside a jacket pocket or backpack, making it difficult for camera systems to determine if a product is being purchased or stolen. What if it were possible to sense what container an item is in, and use this to preemptively prevent theft? 

In this paper, we propose a RFID-based sensing approach, which uses existing RFID tagged items to sense materials. Based on how different materials interact with RFID signals, we can classify the specific material and therefore the container an item is placed in, or even track the container's position with computer vision. Using this material detection system, a cashierless store can sense in real time when items are at risk of theft, and deploy security protocols or have staff intervene. Our system would allow for continual item/container tracking and risk monitoring, improving item-to-customer association in cashierless stores.

\section{Methodology}
RFID (Radio Frequency Identification) is a technology that utilizes wireless communication between a reader antenna and tags. RFID tags either passively harvest energy from the reader's RF signal, or contain an activate onboard power source \cite{weinstein2005rfid, chawla2007overview}. Our system utilizes UHF (ultra-high frequency) passive RFID tags, which modulate the incoming signal from the reader through backscattering in order to return its encoded data. Apart from the tag's ID and other tag data \cite{sawant2015rfid}, RFID readers are also able to measure signal strength with RSSI (Received Signal Strength Indicator), and phase angle, which represents the position in the waveform of the received signal relative to the transmitted signal. 

\subsection{Material Detection}
When backscattered signals from RFID tags pass through materials, the signal can be absorbed, reflected, or scattered because of the properties of the material \cite{chen2024wireless, pradhan2017konark, 5751704}. Depending on the density of the material, attenuation of the RFID signal can occur when RF signals interact with molecules, decreasing the RSSI as measured by the reader. Additionally, materials can scatter RF waves, creating multipath signals that arrive at the reader at different times, causing varying phase shifts \cite{shi2023multipath}. This can lead to constructive or destructive interference, which can increase/decrease the mean RSSI, and also increases the variance in RSSI and phase angle. Lastly, the dielectric constant of the material can affect the propagation speed of RF signals passing through it, leading to a measurable difference in phase shift at the reader \cite{meng2016rfid}. Using these three effects, we have four features that we can use to analyze materials: RSSI mean, RSSI variance, phase angle mean, and phase angle variance. These features can then be fed into a classifier model to determine what material the signal passed through \cite{wang2017tagscan}. A potential concern with using RSSI and phase variance for material detection is that a sampling period is required to collect these features. While many RFID readers can interrogate over a hundred tags per second, a customer would be required to stand within the scanning area of the reader for the entire sampling period to obtain consistent results. In our study, we experimented with single data point samples, which represent one collected measurement with two features for RSSI and phase angle, and one second sampling groups, which have all four features mentioned previously. We determined that a one-second sampling period would be consistent with a customer passing an RFID reader at a reasonable indoor walking pace and that the RFID tag's distance to the reader would only shift slightly.

\subsection{Accounting for Distance}
A major challenge with using RFID RSSI for material sensing is the nonlinear relationship between RSSI and distance \cite{chen2010mechanisms, parameswaran2009rssi}. In order to accurately measure how a material affects RF signal propagation, customers must pass through a point with a known distance, or the distance to the target must be calculated on the fly. In our study, both approaches are explored. We envision a system where RFID readers are placed at strategic points between sections of the store, where customers will pass through while shopping. At these points, our system will detect if items are inside a suspicious container and update theft risk data. It is realistic to assume that in a retail environment, a customer could pass through a location with a fixed distance from the reader, such as between two shelves or a corridor. This idea is explored in our large dataset model. However, in a more open environment, customers may not remain a consistent distance away from readers. In this scenario, a distance sensor mounted close to the antenna, such as an ultrasonic sensor or laser rangefinder, could aid the system by providing a distance measurement. For our experiment, we used various predetermined measured distances to represent on the fly distance measurements. Another challenge is normalizing RSSI for distance. While prior studies have explored formulas for this relationship \cite{bekkali2007rfid, montaser2014rfid}, such as the log distance path loss model, variables such as the model of RFID reader, RFID tag, and RF environment noise can have a significant effect on the model \cite{henriques2014using}. In our experiment, we instead chose to feed the distance as a feature into a neural network model, which is able to capture the nuances in this relationship \cite{mahfouz2015non, hoang2019recurrent, weerasinghe2019rssi, durtschi2024overview}, instead of attempting to normalize the RSSI values beforehand.

\section{Experiment}
We conducted our data collection in a simulated retail environment with seven classes of container commonly found in retail settings. These included a control (no container), a plastic box, a cardboard box, a thin plastic shopping bag, a jacket pocket, a thick fabric bag, and a backpack. We focused on three primary experiments, as outlined in our methodology: a two feature single data point experiment, a four feature 1 second long sample experiment, and a five feature 1 second long sample experiment with varying distances. 
\subsection{Store Setup}
\begin{figure}[htbp]
  \centering
  \includegraphics[width=\linewidth]{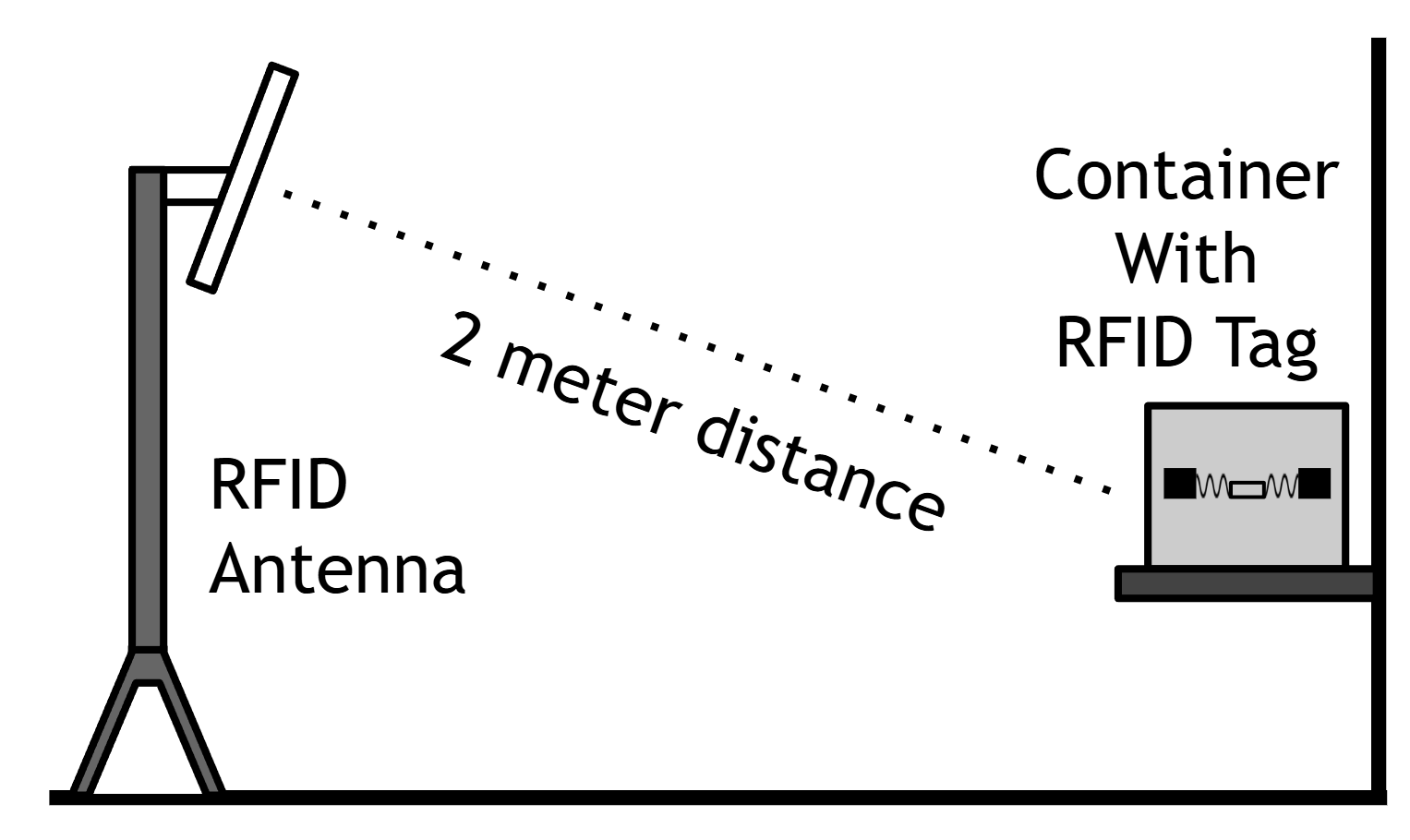}
  \caption{Side view of items in simulated store}
  \Description{Side view of items in simulated store}
  \label{fig:store_side}
\end{figure}

\begin{figure}[htbp]
  \centering
  \includegraphics[width=\linewidth]{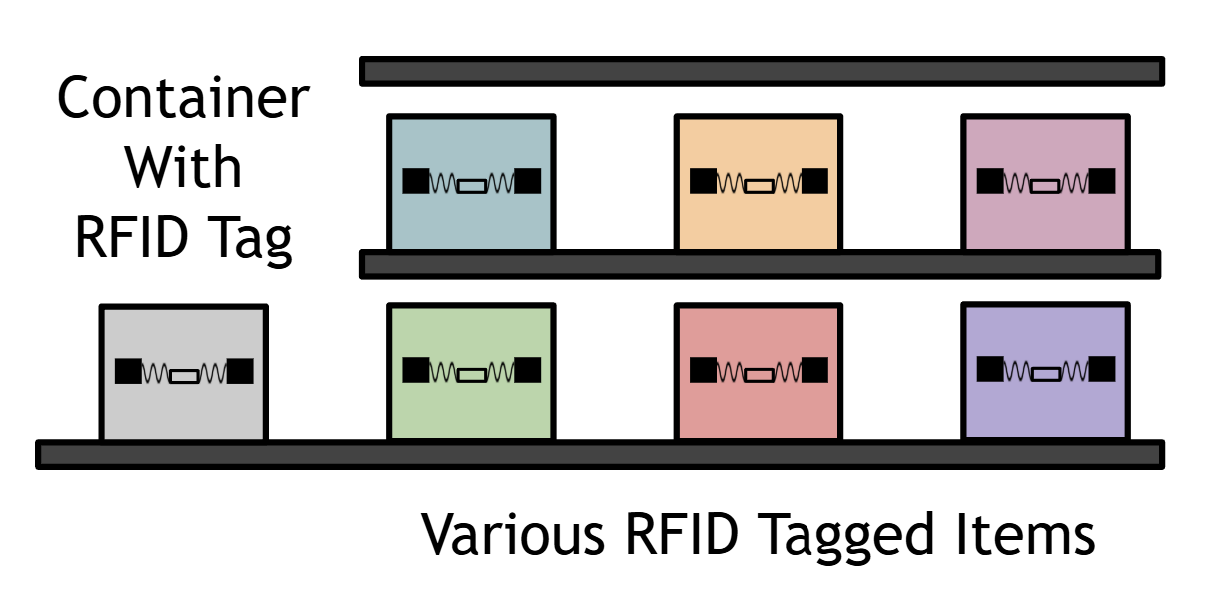}
  \caption{Front view of items in simulated store}
  \Description{Front view of items in simulated store}
  \label{fig:store_front}
\end{figure}

Our simulated retail store setup featured the following configuration, show in Figure~\ref{fig:store_side} and Figure~\ref{fig:store_front}.
We used an Impinj Speedway R420 reader to record RSSI and phase angle data. For all experiments, the RFID reader was configured to the maximum transmit power of 32.5 dBm, and the receive sensitivity was set to the maximum of -84 dBm. In each experiment, a UHF RFID tag with a known ID was placed inside the container and oriented flat with the ground. To replicate a realistic RFID environment, around 55 unique RFID tagged items were placed on the shelves surrounding the data collection point. We found that under these conditions, it was possible for our sampling tag to be interrogated by the reader five times per second. Data collection was split into two phases: initially we collected data continually for 10 minutes at a distance of 2 meters for each of the 7 containers, then collected data continually for 2 minutes at 30 centimeters and 1 meter for the distance correlated samples. Data was either treated as single point samples, or divided into 1 second long samples for the sampling period experiment and the distance experiment. This resulted in around 3000 samples per class in the single point experiment, 600 samples per class for the sampling period experiment, and 120 samples per class for the distance experiment. For the distance experiment, we utilized only the first 2 minutes of data collected from the 2 meter 10 minute tests in order to maintain a balanced dataset for our classifier \cite{buda2018systematic}.

\subsection{Machine Learning Classifier}
For this study, we considered using either a Support Vector Machine (SVM) with a radial basis function kernel or a feedforward neural network. In our initial trial, which consisted of 15-second-long single data point samples from 5 classes of containers, we found that after tuning hyperparameters, the performance of both techniques was nearly identical, with less than a 1\% difference between them. For the actual experiment, due to the large sample sizes (around ~20000 individual data points), we chose to use the neural network as the training time was much lower than that of the SVM. We partitioned the dataset into 70\% training, 15\% validation, and 15\% testing sets. To train the model, we used the popular Python machine learning libraries Keras, Tensorflow, and Scikit Learn, and incorporated automatic hyperparameter turning with the Keras Tuner library. Specific details about the neural network architectures used are shown in the Results section. For all three experiments, we used ReLU as the hidden layer activation function, Softmax as the output layer activation function, Adam as the optimizer, and Categorical Crossentropy as our loss function. 

\subsection{Limitations}
In our study, we did not use RFID tags attached to store items, instead measuring only tags themselves. Due near field coupling between the antenna and the item the tag is attached to, changes in antenna impedance may alter RSSI \cite{siden2007remote, gonccalves2015rfid}. We believe that this effect can be measured for different items, and the tag ID could be used to look up a correction factor to apply when the item is detected, but this needs to be explored further. Additionally, we did not account for people passing between the antenna and scanned items, something explored in other studies \cite{9219250, li2015idsense}, which may have an effect on RFID similar to that of a container.

\section{Results}
\subsection{Raw Data}
\begin{figure}[htbp]
    \centering
    \begin{subfigure}[t]{7.5cm}
        \centering
        \includegraphics[width=\linewidth]{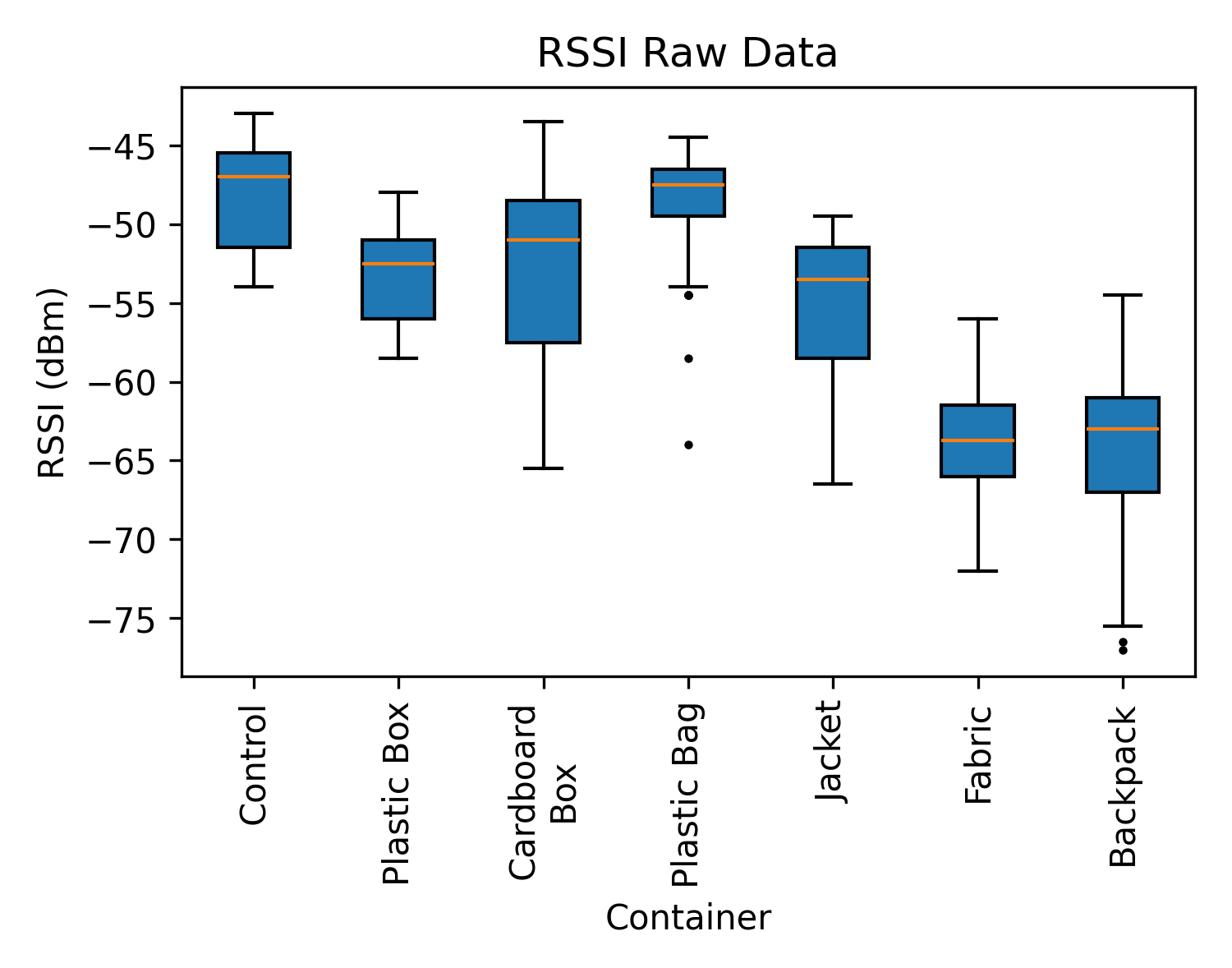}
        \caption{Observed RSSI}
    \end{subfigure}
    \hfill
    \begin{subfigure}[t]{7cm}
        \centering
        \includegraphics[width=\linewidth]{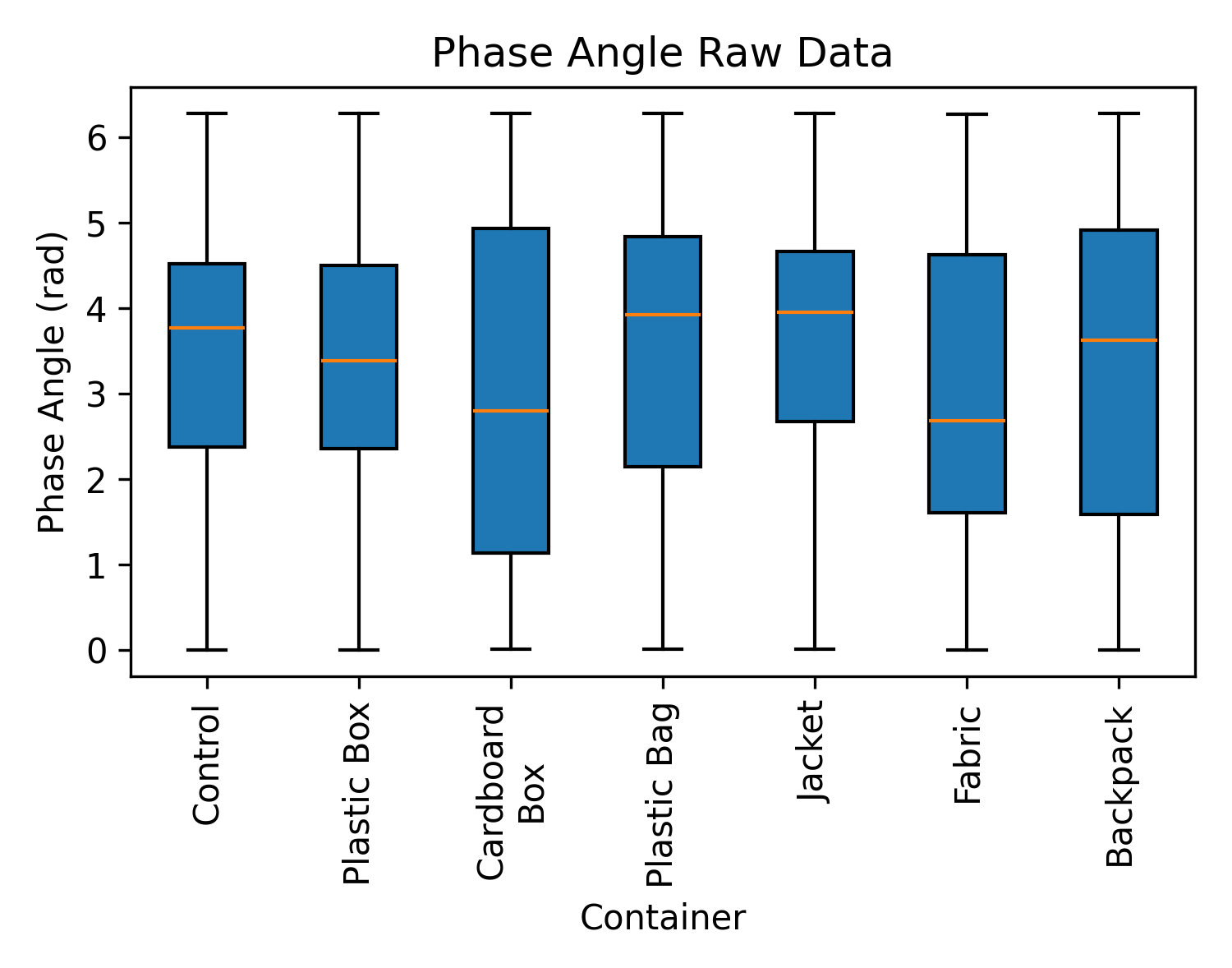}
        \caption{Observed Phase Angle}
    \end{subfigure}
    \caption{Raw Data}
    \label{fig:raw}
\end{figure}
We observed the RSSI and phase angle shown in Figure~\ref{fig:raw} data during the 10 minute 2 meter data collection periods. As shown, the distributions for the 7 classes are visually distinct in the box plots. Consistent with what was discussed in our RF material sensing methodology, thinner materials, such as the plastic bag, plastic box, and cardboard box, absorb less RF signal compared to thicker materials like the fabric bag and backpack. Additionally, complex multilayer materials, such as the cardboard box and backpack, had a larger variance in RSSI and phase angle, explained through scattering caused by interactions with multiple layers. 
\subsection{Model Tuning}
We utilized a combination of automated hyperparameter turning (Samples w/ Distance) and manual turning (Single Data Point and One Second Samples) to achieve better performance. The model architectures shown in Table~\ref{tab:Architecture} were used, which had the highest accuracy.
\begin{table}[htbp]
    \centering
    \begin{tabular}{p{1.75cm}|p{1.75cm}p{1.75cm}p{1.75cm}}
        \toprule
        \textbf{ } & \textbf{Single Data Points} & \textbf{One Second Samples} & \textbf{Samples w/ Distance} \\
        \midrule
        Features & 2 & 4 & 5 \\
        Classes & 7 & 7 & 4 \\
        Batch Size & 16 & 16 & 16 \\
        Learning Rate & 0.001 & 0.001 & 0.001 \\
        Epochs & 115 & 105 & 55 \\
        Parameters & 11,367 & 11,207 & 1,396\\
        Neurons Per Hidden Layer & 128, 64, 32, 16 & 128, 64, 32 & 32, 16, 32 \\
        \bottomrule
    \end{tabular}
    \caption{Neural Network Architecture}
    \Description{Neural Network Architecture}
    \label{tab:Architecture}
\end{table}

\begin{figure}[htbp]
    \centering
    \begin{subfigure}[t]{4cm}
        \centering
        \includegraphics[width=\linewidth]{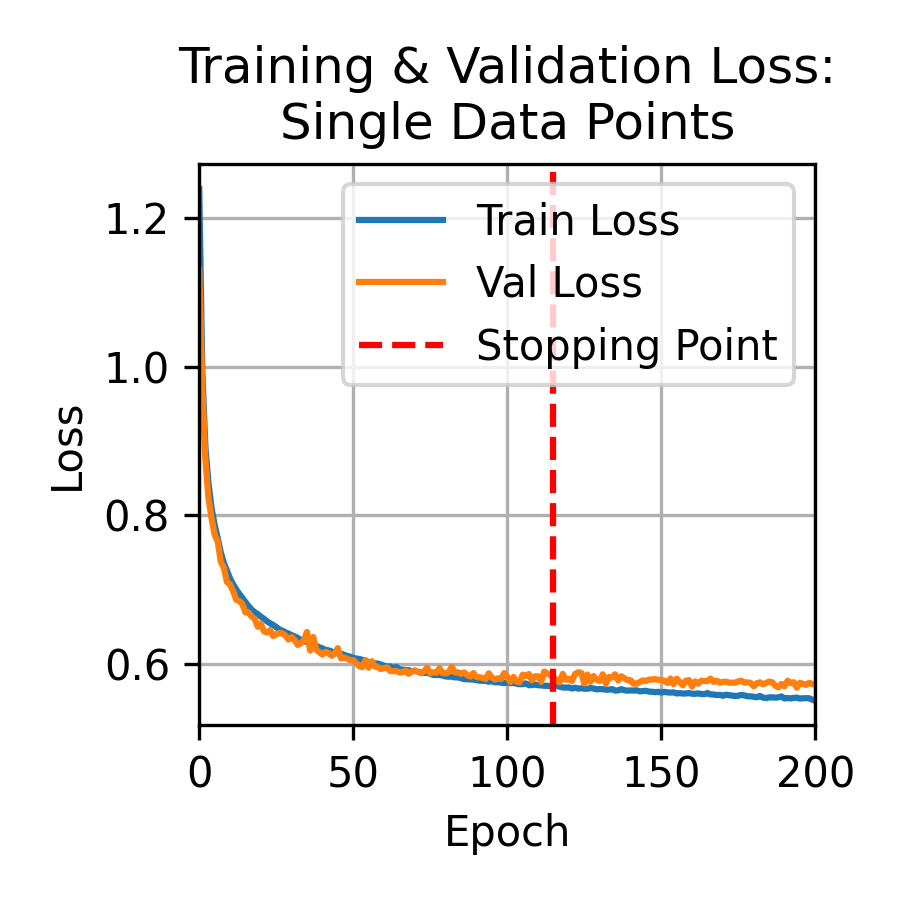}
        \caption{Single Data Points}
    \end{subfigure}
    \hfill
    \begin{subfigure}[t]{4cm}
        \centering
        \includegraphics[width=\linewidth]{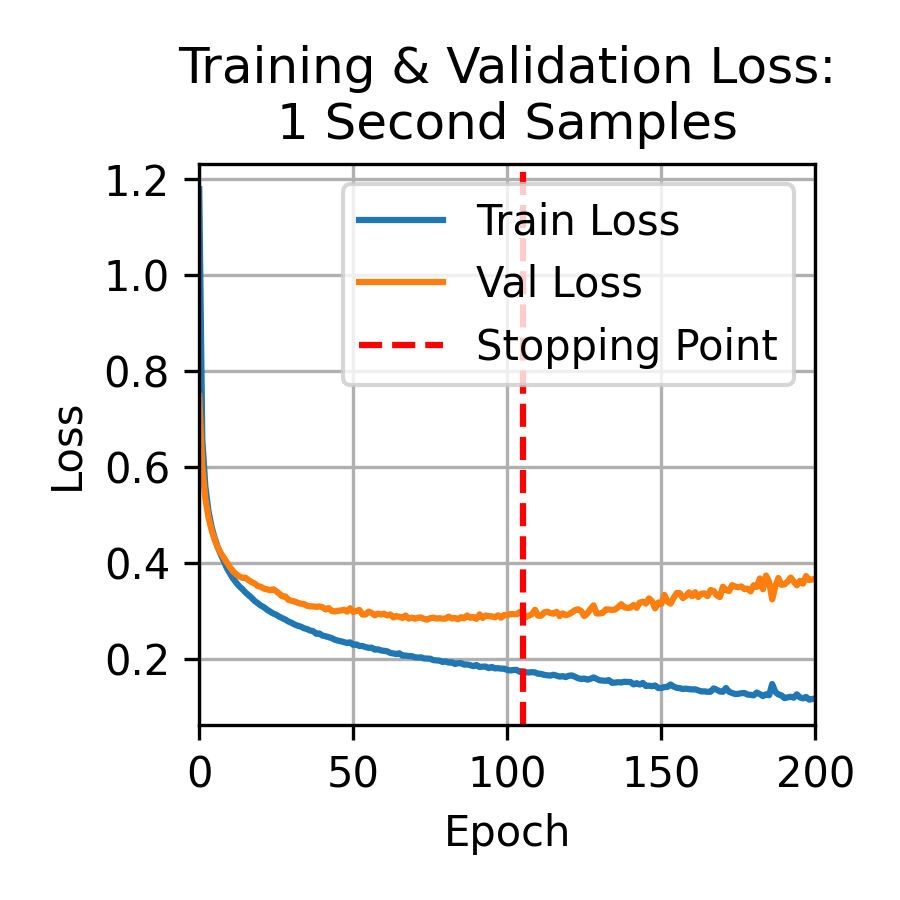}
        \caption{1 Second Samples}
    \end{subfigure}
    \begin{subfigure}[t]{4cm}
        \centering
        \includegraphics[width=\linewidth]{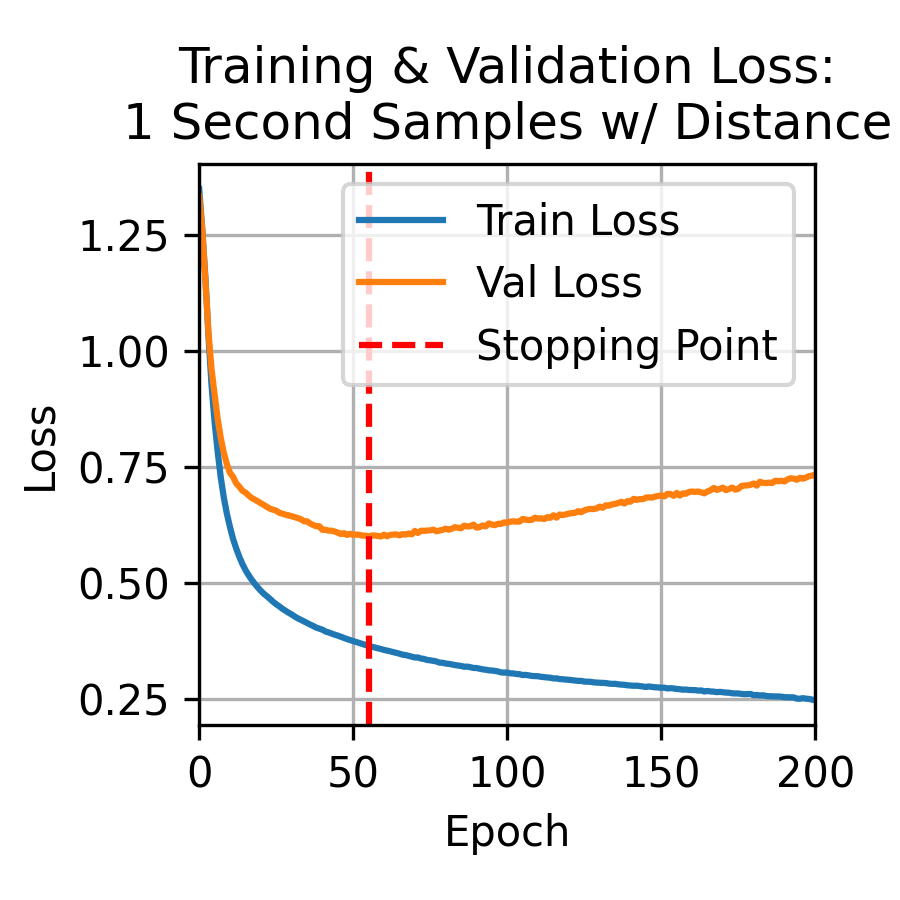}
        \caption{1 Second Samples w/ Distance}
    \end{subfigure}
    \caption{Training and Validation Loss}
    \label{fig:tv}
\end{figure}

We utilized early stopping to prevent overfitting in the neural networks for each experiment. Figure~\ref{fig:tv} contains the training/validation loss plots for each experiment, with the stopping points marked.

\begin{figure}[htbp]
  \centering
  \includegraphics[width=\linewidth]{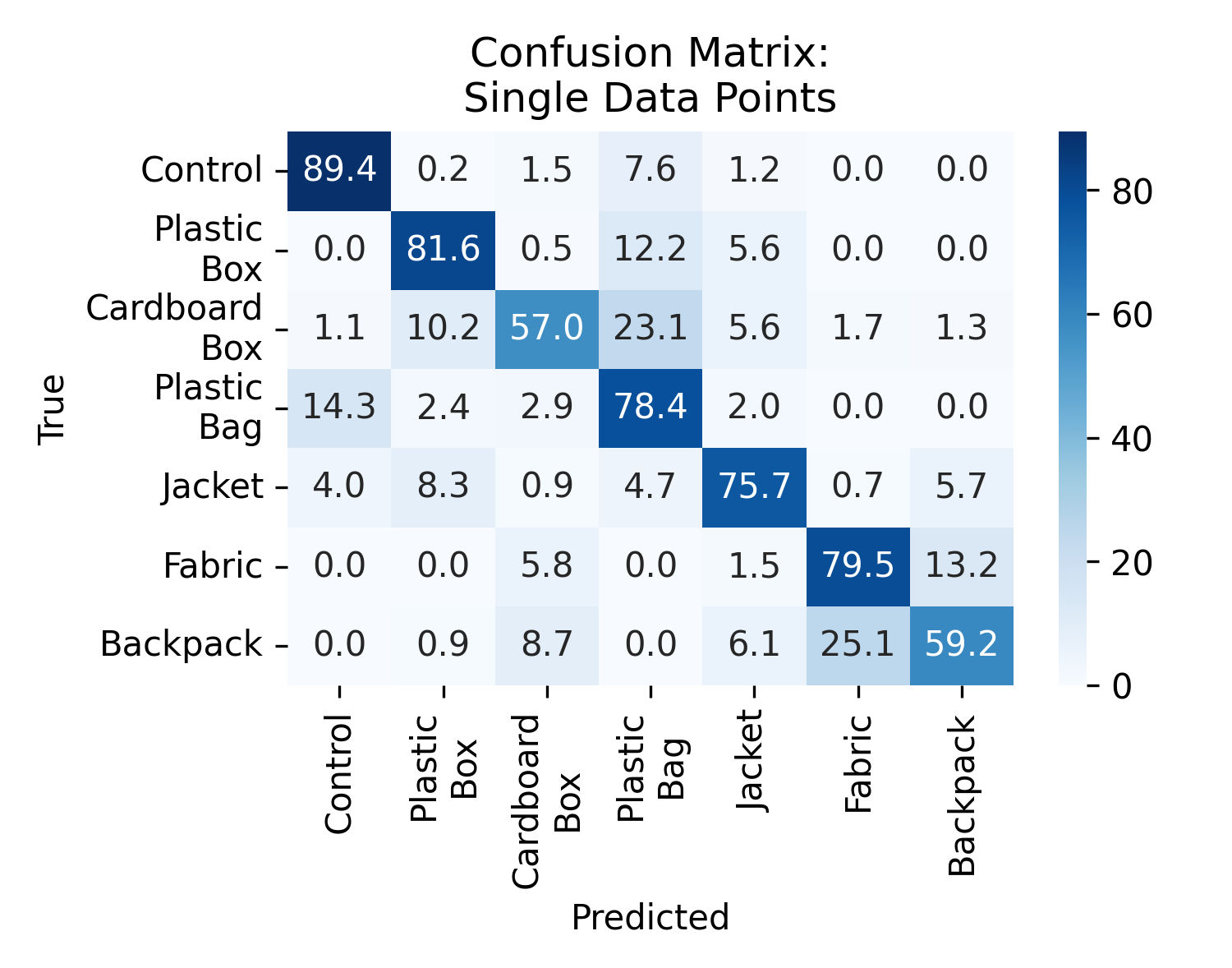}
  \caption{Single Data Points Confusion Matrix}
  \Description{Single Data Points Confusion Matrix}
  \label{fig:single}
\end{figure}

\begin{figure}[htbp]
  \centering
  \includegraphics[width=\linewidth]{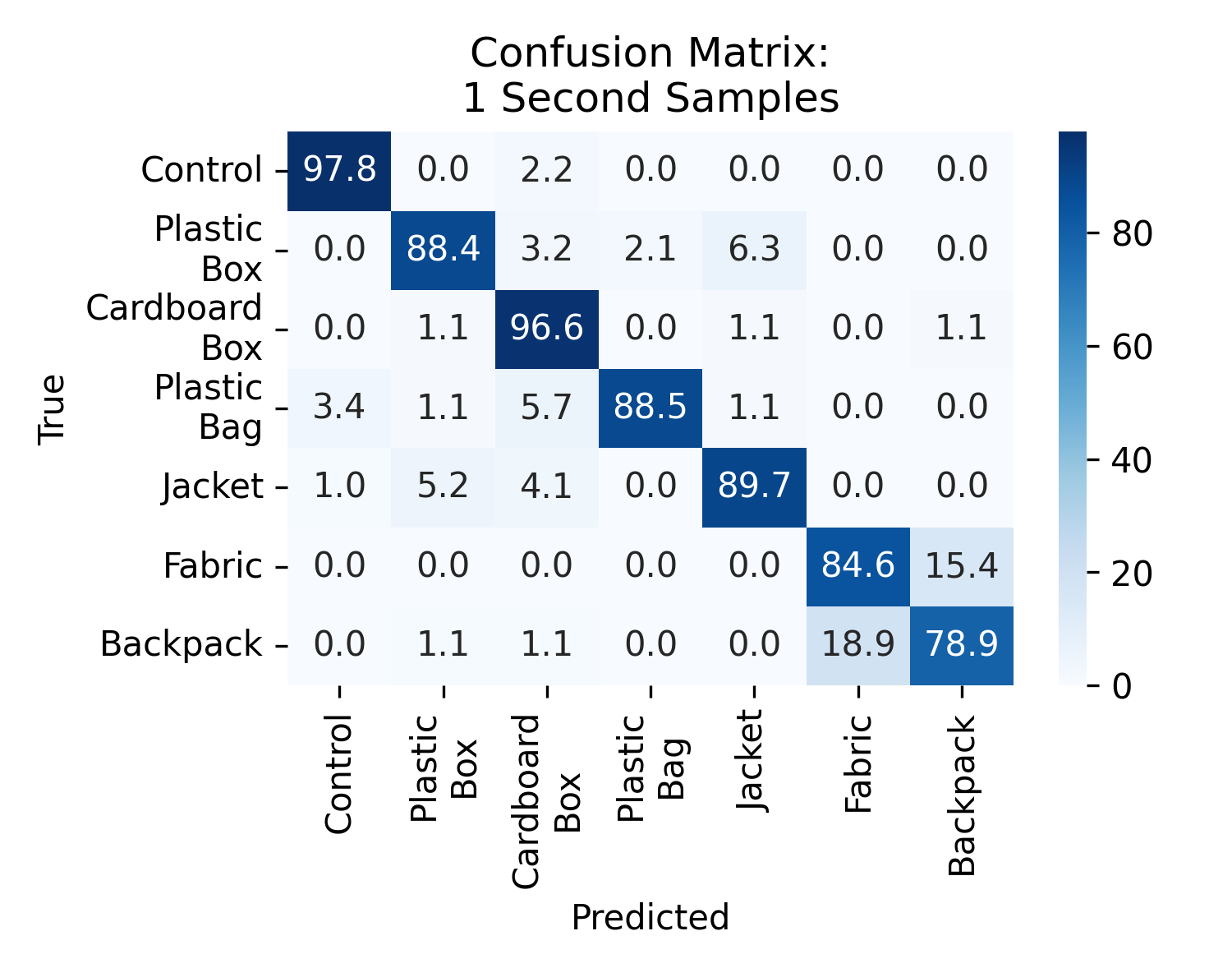}
  \caption{1 Second Samples Confusion Matrix}
  \Description{1 Second Samples Confusion Matrix}
  \label{fig:sample}
\end{figure}

\begin{figure}[htbp]
  \centering
  \includegraphics[width=\linewidth]{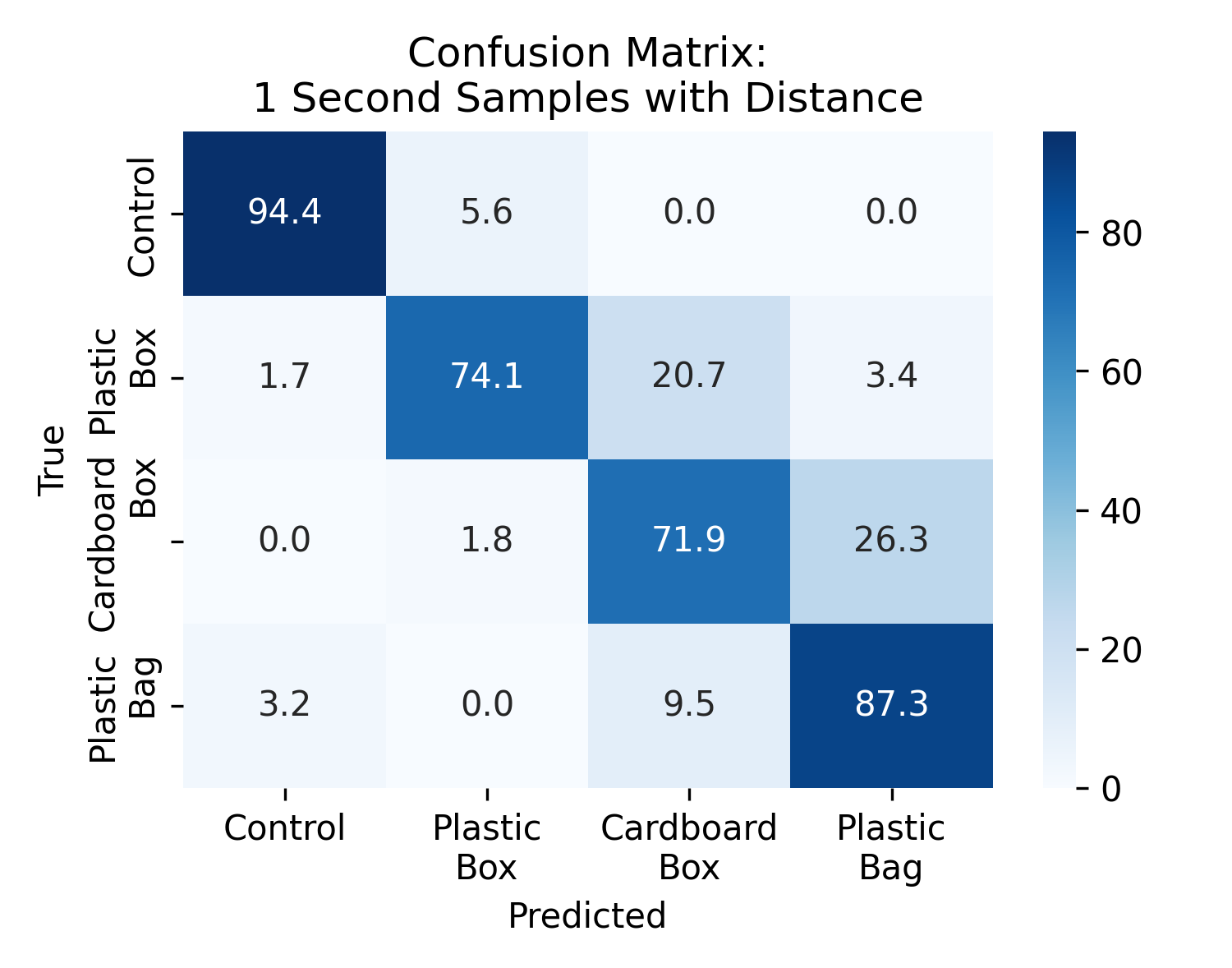}
  \caption{Samples With Distance Confusion Matrix}
  \Description{Samples With Distance Confusion Matrix}
  \label{fig:dist}
\end{figure}

\subsection{Single Data Points}
For the single data point experiment, shown in Figure~\ref{fig:single}, we recorded a testing accuracy of 0.7412. Due to the lack of variance data, the model struggled more with differentiating materials with higher RF scattering, such as the cardboard box and backpack. We also found that the classifier for this experiment required additional layers compared to the other models, likely due to the weaker association. While we initially believed that the performance would suffer greatly without this data, the results proved more accurate than expected, and could be considered a viable solution in environments where customers may not stay in the same position for very long.

\subsection{One Second Samples}
For the one second sample experiment, shown in Figure~\ref{fig:sample}, we recorded a testing accuracy of 0.8921, the highest out of the three scenarios. This was in line with our expectations, given the addition of variance data, and compared to the single point experiment, the classifier was able to perform much better with the cardboard box. However, we noted some confusion between the fabric and backpack, likely due to the similarities between the two materials.

\subsection{Samples With Distance}
For the samples with distance experiment, shown in Figure~\ref{fig:dist}, we recorded a testing accuracy of 0.8190. Given the challenges with differentiating RSSI at different distances, we were surprised by the relatively high accuracy, indicating that the addition of distance measurements to the classifier inputs is able to compensate for the nonlinear behavior of RSSI at different distances.

\section{Conclusion}
Our experiments show that RFID features can identify the container around a tagged item with up to 89\% accuracy, providing reliable material identification with standard RFID tags. When placed at pinch points, our system can flag hidden items in backpacks, pockets, or bags, prompting camera checks or staff alerts, greatly improving theft detection. Because this method relies on existing RFID infrastructure and requires only a short sample for high accuracy, it is both low cost and real time. Future work should investigate item specific tag coupling, test performance with human occlusion, and evaluate the approach in a live store. By converting RFID tags into sensors for container detection, this technique gives cashierless retail proactive and unobtrusive loss prevention.

\bibliographystyle{ACM-Reference-Format}
\bibliography{references}

\end{document}